\documentclass[aps,prl,twocolumn,preprintnumbers,superscriptaddress]{revtex4}

\usepackage{graphicx}
\usepackage{tabularx}
\usepackage{amssymb}
\usepackage{amsmath}
\usepackage{bm}
\usepackage{subfigure}
\usepackage{graphpap}
\usepackage{multirow}
\usepackage{notes2bib}
\usepackage[version=3]{mhchem} 
\begin{document}
\title{Hybrid Organic-Inorganic Perovskites as Promising Substrates for Pt Single-Atom Catalysts}

\author{Qiang Fu}
\email{qiang@physik.hu-berlin.de; qfu@sdu.edu.cn}
\affiliation{Institut f$\ddot{u}$r Physik and IRIS Adlershof, Humboldt-Universit$\ddot{a}$t zu Berlin, 12489 Berlin, Germany}
\affiliation{School of Chemistry and Chemical Engineering, Shandong University, 250100 Jinan, China}
\author{Claudia Draxl}
\affiliation{Institut f$\ddot{u}$r Physik and IRIS Adlershof, Humboldt-Universit$\ddot{a}$t zu Berlin, 12489 Berlin, Germany}
\affiliation{Fritz-Haber-Institut der Max-Planck-Gesellschaft, 14195 Berlin, Germany}

\begin{abstract}
Single-atom catalysts (SACs) combine the best of two worlds by bridging heterogeneous and homogeneous catalysis. The superior catalytic properties of SACs, however, can hardly be exploited without a suitable substrate. Here, we explore the possibility of using hybrid organic-inorganic perovskites as supporting materials for single transition-metal atoms. By means of first-principles calculations, we predict that single Pt atoms can be incorporated into methylammonium lead iodide surfaces by replacing the methylammonium groups at the outermost layer. The iodide anions at the surface provide potentially uniform anchoring sites for the Pt atoms and donate electrons, generating negatively-charged Pt$_{1}^{\delta-}$ species that allow for preferential O$_{2}$ adsorption in the presence of CO. Such Pt sites are able to catalyze CO oxidation and may also play a role in CO$_{2}$ reduction. The fundamental understanding generated here will shed light on potential applications of hybrid perovskites in the field of (photo)catalysis.
\end{abstract}

\maketitle
Platinum as a catalyst plays a central role in energy conversion, pollution remediation, and petrochemical industry. Owing to the scarcity of the element, it is highly desirable to downsize Pt particles as much as possible. Considerable effort has been devoted to synthesizing and dispersing sub-nano Pt clusters \cite{Fu2003,Vajda2009}, whose size can be reduced to the ultimate form of single atoms \cite{ZhaiY2010,Qiao2011,Moses-DeBusk2013,Kistler2014,DingK2015,Dvorak2016,Gao2016,Nie2017,Wei2018,Therrien2018}. While such type of catalysts exhibits superior properties in different reactions, single atoms need to be anchored on support materials \cite{LiuJY2017,Wang2018}. A suitable substrate should be able to prevent aggregation of single atoms but preserve their catalytic activity at the same time. Optimal selectivity of a reaction is expected when the atoms are uniformly distributed on the substrate, where all anchoring sites are \textit{atomically} equivalent \cite{YangD2017}. Depositing single atoms on a photoactive support material offers further strategies for designing photocatalysts \cite{Gao2016,Fang2018}. Despite being a challenge, integrating all the above features in one substrate is of great importance for tailoring single-atom catalysts with extraordinary performance.

Hybrid organic-inorganic perovskites (HOIPs), with the chemical formula ABX$_{3}$, consist of organic cations A$^{+}$ and an inorganic anionic framework BX$_{3}^{-}$ \cite{Walsh2015}. Besides suitable band gaps for harvesting sunlight, they exhibit fascinating properties like high optical absorption, electrically benign dominating defects, high charge-carrier lifetimes, low effective carrier masses, and long diffusion lengths \cite{Yin2014,Colella2016}. Their optoelectronic properties can also be tuned through independent substitutions on the A, B, and X sites \cite{Walsh2015,Yin2014}. Among the various combinations, methylammonium lead iodide (MAPbI$_{3}$, MA$=$CH$_{3}$NH$_{3}$) is the most widely studied HOIP. While most research on HOIPs is aimed at photoexcitations and photovoltaic applications, the materials have recently been demonstrated to exhibit great potential in the field of heterogeneous photocatalysis \cite{ParkS2016,XuYF2017,ChenK2017,WuY2018,HuangH2018}.

Combining single-atom catalysis and hybrid organic-inorganic perovskites may bring about novel catalytic systems with extraordinary performance. Not only the presence of transition-metal SACs can effectively enhance the interaction between HOIPs and reactant molecules, thereby promoting the activation of molecular chemical bonds; also, HOIP substrates could induce distinct catalytic properties to the transition-metal SACs. Moreover, the unique lattice structure for providing regular surface patterns and the outstanding ability of absorbing sunlight also call for employing HOIPs like MAPbI$_{3}$ as a potential substrate for SACs. In the present work, we investigate this possibility through a first-principles study. We focus on CO oxidation, a seemingly simple but prototypical reaction \cite{Ertl1994,Freund2011}, to examine the corresponding catalytic properties. Such process can take place at room temperature \cite{Newton2015,Gatla2016} and does not involve water molecules that may deteriorate the stability of the MAPbI$_{3}$ lattice \cite{Christians2015,Mosconi2015,Zhang2016,Yang2017}. The MAI-terminated (110) and (001) surfaces of tetragonal MAPbI$_{3}$ are adopted as the supporting substrates. Such type of termination, which was theoretically predicted to be much more stable than that of PbI$_{2}$ \cite{Quarti2017,GengW2015}, has been experimentally verified by scanning tunneling microscopy  \cite{SheL2016,Ohmann2015}. We find that single Pt atoms can adsorb on MAPbI$_{3}$ by replacing surface MA groups, while it is thermodynamically unfavorable for single atoms of Cu, Ag, Au, and Pd. Here, each Pt atom is bound to an anchor site that is composed by four I$^{-}$ anions. Charge transfer from the anionic PbI$_{3}^{-}$ framework makes the Pt$_{1}$ species negatively charged, in contrast to the cationic Pt$_{1}^{\delta+}$ active sites that were consistently reported in previous studies \cite{ZhaiY2010,Qiao2011,Moses-DeBusk2013,Kistler2014,DingK2015,Dvorak2016,Gao2016,Nie2017,Wei2018}. Importantly, we find that O$_{2}$ can preferentially occupy the Pt sites and become activated, even though CO molecules coexist. By demonstrating the feasibility of catalytic CO oxidation, our work sheds light on potential applications of hybrid perovskites in (photo)catalysis.

The substrates were simulated using slab models, consisting of four layers of the MAPbI$_{3}$ unit and being terminated by MA groups at the top. Each layer has four MA$^{+}$ cations, corresponding to a 2$\times$2 unit cell for the (110) orientation and a $\sqrt{2}\times\sqrt{2}$-R45$^{\circ}$ unit cell for the (001) one. As such, the surface unit cell contains 192 atoms. The bulk lattice parameters (a $=$ 8.800 {\AA}, c $=$ 12.685 {\AA}) were taken from experiments by Kawamura and coworkers \cite{Kawamura2002}. All calculations were performed using spin-polarized density functional theory (DFT) as implemented in VASP \cite{vasp1,vasp2}, employing the projector augmented wave (PAW) method \cite{PAW}. A plane-wave energy cutoff of 400 eV was used. Exchange-correlation interactions were described by the optB86b-vdW functional \cite{Dion04,Klimes11}, which was found to perform very well in similar systems \cite{WangY2014}. The Brillouin zone (BZ) was sampled using a 3$\times$3$\times$1 Monkhorst$-$Pack grid \cite{MP}. To eliminate spurious interactions between images, a 16-{\AA} vacuum region was added in the vertical direction, and a dipole correction was applied to the total energies. The atomic coordinates in the top two formula units were relaxed until the maximum residual force was less than 0.03 eV/{\AA}. The transition states were located by using the climbing-image nudged elastic band method \cite{CINEB} and the dimer method \cite{Dimer} with a force criterion of 0.05 eV/{\AA} for atomic relaxations.

\begin{table}[b!]
\centering
\caption{Adsorption energies (in eV) of the single transition-metal atoms of Cu, Pd, Ag, Pt, and Au as well as MA on the MAPbI$_{3}$ (110) and (001) substrates.}
\scalebox{1.15}{\begin{tabular}{c|c|c|c|c|c|c}
\hline
 Orientation &   Cu   &   Pd   &   Ag   &   Pt   &   Au   &   MA   \\
\hline
     (110)   & -4.25  & -3.87  & -3.61  & -5.23  & -3.74  & -4.56  \\
	   (001)   & -4.06  & -3.65  & -3.38  & -4.98  & -3.58  & -4.25  \\
\hline
\end{tabular}}
\label{ad-M1}
\end{table}

In the bulk, MA$^{+}$ cations are confined within the anionic framework by a strong electrostatic potential \cite{Walsh2015}. At the surface layer, by contrast, the negatively charged PbI$_{3}^{-}$ cages are open and as such, the MA species may be replaced. To investigate this possibility, we consider the replacement of MA by a single atom of transition-metals, TM$_{1}$ (TM = Cu, Pd, Ag, Pt, and Au). It is worth noting that the MA group that is released from the cage is a neutral radical, as the total number of electrons in the substrate has to be conserved. The interaction between TM$_{1}$/MA and the MAPbI$_{3}$ host, which lacks one surface MA, is examined. The adsorption energy,
\begin{equation}
  \label{eq:Ads-M1}
  E_{ads} = E_{host-TM_{1}} - (E_{host} + E_{TM_{1}}),
\end{equation}
is used to describe the interaction strength. Here, $E_{host}$ and $E_{TM_{1}}$ are the energies of the host and the isolated TM$_{1}$ species, respectively, and $E_{host-TM_{1}}$ represents the energy of the combined system with TM$_{1}$ deposited on the surface of the host. A negative $E_{ads}$ value corresponds to exothermic adsorption. The adsorption energies of TM$_{1}$/MA on both the (110) and the (001) surfaces are listed in Table \ref{ad-M1}. Interestingly, we find that among the five metal elements, only Pt exhibits a higher adsorption energy than MA. This result demonstrates that one can indeed implant single Pt atoms into the surface layer of the substrate. A more detailed discussion on the stability of the anchored Pt$_{1}$ species can be found in the Supplemental Material \cite{SM}. Moreover, since the same conclusion holds for both surfaces, such implantation is expected to be independent on a particular orientation of MAPbI$_{3}$.

\begin{figure}[t!]
\centering
  \includegraphics[width=8.0cm]{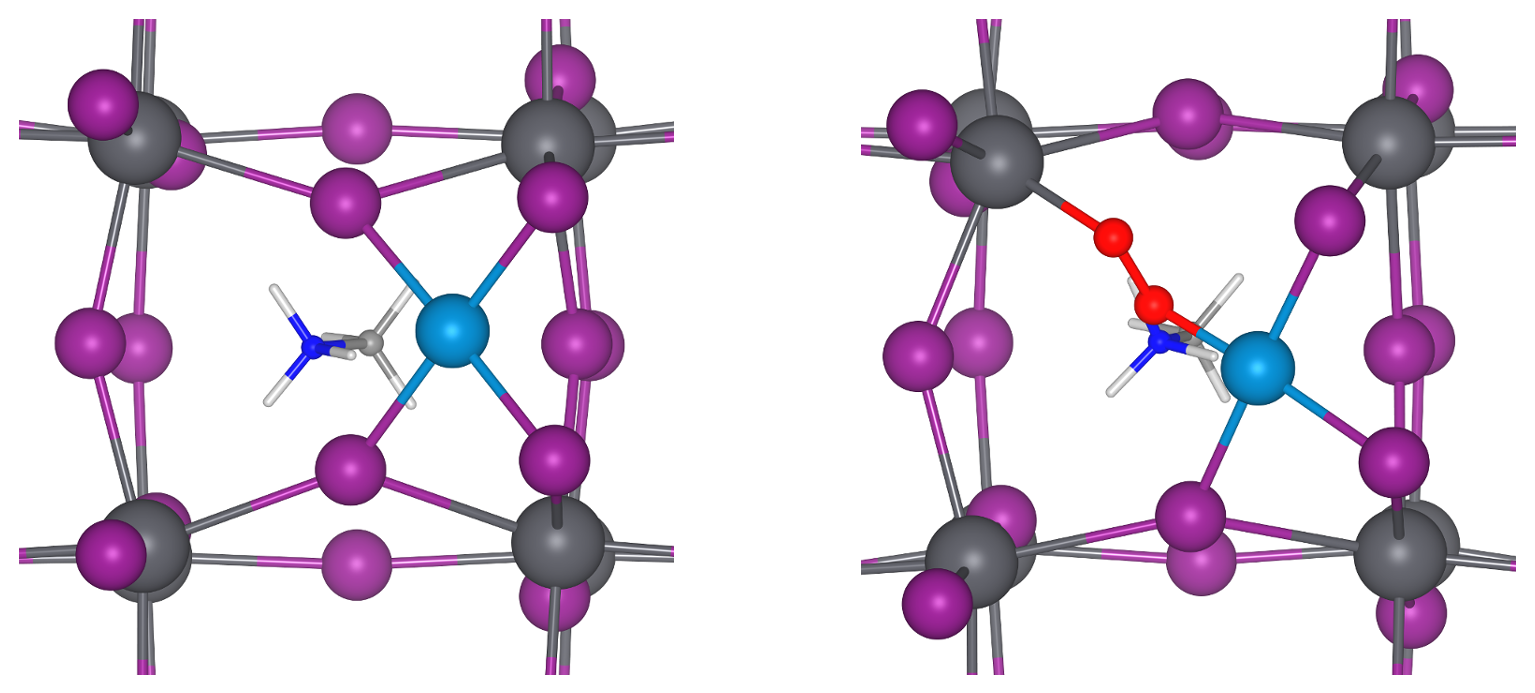}
  \caption{Geometries of the Pt$_{1}$@(110) system before (left) and after (right) O$_{2}$ adsorption. Purple and dark gray spheres indicate the iodine and lead atoms of the cage, and steel blue and red spheres represent the platinum and oxygen atoms. The MA group is displayed as a stick model with carbon, nitrogen, and hydrogen atoms depicted in light gray, blue, and white, respectively.}
  \label{Model}
\end{figure}

Upon deposition on the (110) substrate, the single Pt atom is bonded to four iodine anions, as displayed in the left panel of Fig. \ref{Model}. This configuration could provide uniform anchoring sites for the potential single-atom catalyst. Here, the Pt single atom forms covalent bonds with the four neighboring iodine anions, which is demonstrated by the density of states (DOS) in which the Pt \textit{d} orbitals substantially overlap with those of the iodine \textit{p} orbitals (Fig. S1). Bader charge analysis reveals that the Pt atom carries a small amount of negative charge of 0.13 electrons, while the number of negative charges carried by the surrounding iodine anions decreases from 0.51 to 0.36 upon the deposition (Fig. S2). The anionic PbI$_{3}^{-}$ framework, or more precisely, the nearby iodine anions, share their extra electrons with Pt. Such electron transfer reduces the overall electrostatic repulsion and further increases the stability of the system (Fig. S3). Since the iodine species here are intrinsically negatively charged, the aforementioned charge transfer toward Pt is expected to be general and does not depend on a specific pattern of the Pt$-$I bonds. Calculations considering deposited Pt$_{2}$ and Pt$_{3}$ clusters (Figs. S4 and S5) confirm this supposition. Bader charge analysis demonstrates that the two sub-nano clusters are also negatively charged, obtaining 0.47 and 0.70 electrons from the substrate, respectively.

The single Pt atom behaves as a potential adsorption site for reactant molecules like O$_{2}$, as demonstrated in the following. In the right panel of Fig. \ref{Model}, we present the geometry of an adsorbed O$_{2}$ molecule on the Pt$_{1}$@(110) system. Here, one of the two O atoms is bonded to the Pt atom, whereas the other one connects with a Pb atom nearby. Upon adsorption, the O$_{2}$ molecule becomes activated, exhibiting an increased O$-$O bond length that is 0.11 {\AA} longer than that of an isolated molecule. Such activation originates from charge transfer to the adsorbate. Bader charge analysis reveals that O$_{2}$ obtains from the Pt atom and I atoms nearby as much as 0.64 electrons, which occupy the anti-bonding $\pi^{*}$ orbital of the molecule and thereby, weaken the O$-$O bond. For comparison, the O$_{2}$ adsorption is also investigated at the Pt$_{2}$@(110), Pt$_{3}$@(110), and Pt$_{1}$@(001) systems (Figs. S4-S6). In Table \ref{data-O2}, the O$-$O bond lengths and the amounts of charge carried by O$_{2}$ are listed. One can see that in all the three cases, the O$-$O bond lengths are stretched, and electrons are transferred to the O$_{2}$ adsorbate, both manifesting O$_{2}$ activation. The results indicate that the activation of adsorbed O$_{2}$ is a general feature of Pt$_{n}$@MAPbI$_{3}$ systems, where Pt behaves as a \textquotedblleft bridge\textquotedblright\ for the charge transfer toward the adsorbate.

\begin{table}[t!]
\centering
\caption{O$-$O bond length and amount of charge carried by the O$_{2}$ adsorbate.}
\scalebox{1.15}{\begin{tabular}{c|c|c}
\hline
           System            & Length ({\AA}) & Charge (e$^{-}$) \\
\hline
	 O$_{2}$/Pt$_{1}$@(110)    &      1.35      &        0.64      \\
	 O$_{2}$/Pt$_{2}$@(110)    &      1.40      &        0.59      \\
	 O$_{2}$/Pt$_{3}$@(110)    &      1.39      &        0.59      \\
	 O$_{2}$/Pt$_{1}$@(001)    &      1.35      &        0.68      \\
	   O$_{2}$ (isolated)      &      1.24      &        0.00      \\
\hline
\end{tabular}}
\label{data-O2}
\end{table}

\begin{table}
\centering
\caption{Adsorption energies of CO and molecular O$_{2}$ on Pt$_{1}$@MAPbI$_{3}$. E$_{ad}$(CO/CO) and E$_{ad}$(O$_{2}$/CO) represent the consecutive
adsorption energies of CO and O$_{2}$, respectively, when another CO molecule has been adsorbed already.}
\scalebox{1.15}{\begin{tabular}{l|c|c}
\hline
        Energy (eV)      & Pt$_{1}$@(110) &  Pt$_{1}$@(001) \\
\hline
	     E$_{ad}$(CO)      &     -1.68      &      -1.60      \\
	   E$_{ad}$(O$_{2}$)   &     -1.41      &      -1.27      \\
	    E$_{ad}$(CO/CO)    &     -0.87      &      -0.90      \\
	 E$_{ad}$(O$_{2}$/CO)  &     -1.12      &      -0.79      \\
\hline
\end{tabular}}
\label{Eds}
\end{table}

It is known that on Pt surfaces, the adsorption of CO is much stronger than that of molecular O$_{2}$. As a result, the Pt sites are preferentially accommodated by CO, making O$_{2}$ adsorption remarkably inhibited. Regarding deposited single Pt atoms, there is still no consensus on their catalytic activity toward CO oxidation \cite{Qiao2011,Moses-DeBusk2013,DingK2015}. While Pt$_{1}$ species were demonstrated to be the active sites \cite{Qiao2011,Moses-DeBusk2013}, Ding and coworkers found a lack of activity that was attributed to strong CO binding on Pt$_{1}$ \cite{DingK2015}. As for the Pt$_{1}$ system investigated here, the Pt sites might be blocked by CO and thus would not play a role for the O$_{2}$ activation. To address this question, we calculate the adsorption energies of CO and O$_{2}$, with results listed in Table \ref{Eds}. Both the initial and the consecutive adsorption energies are calculated and compared. For the Pt$_{1}$@(110) and the Pt$_{1}$@(001) systems, the initial adsorption energies of CO are 0.27 eV and 0.33 eV larger, respectively, than the adsorption energies of O$_{2}$. It should be noted, however, that such differences are much smaller than the corresponding value on the Pt(111) surface, which is above 1 eV \cite{Yeo1997,Eichler1997,Fu2011}. The results indicate that the inhibition of O$_{2}$ adsorption by CO is remarkably weakened, which is consistent with recent experiments on other single-atom systems \cite{Peterson2014,LiuJL2016}. Furthermore, tackling this inhibition becomes particularly effective when a CO molecule is already adsorbed at Pt$_{1}$. One can see that at Pt$_{1}$@(001), the consecutive adsorption energy of CO is only 0.11 eV larger than that of O$_{2}$; whereas on the Pt$_{1}$@(110) system, it is the additional O$_{2}$ molecule, rather than CO, that preferentially adsorbs at the Pt site. Hence, we demonstrate that on Pt$_{1}$@MAPbI$_{3}$, the O$_{2}$ adsorption as well as its subsequent activation cannot be blocked by CO reactants. In what follows, we focus on the Pt$_{1}$@(110) system for describing the detailed process of the catalytic CO oxidation.

\begin{figure}[t!]
\centering
  \includegraphics[width=8.5cm]{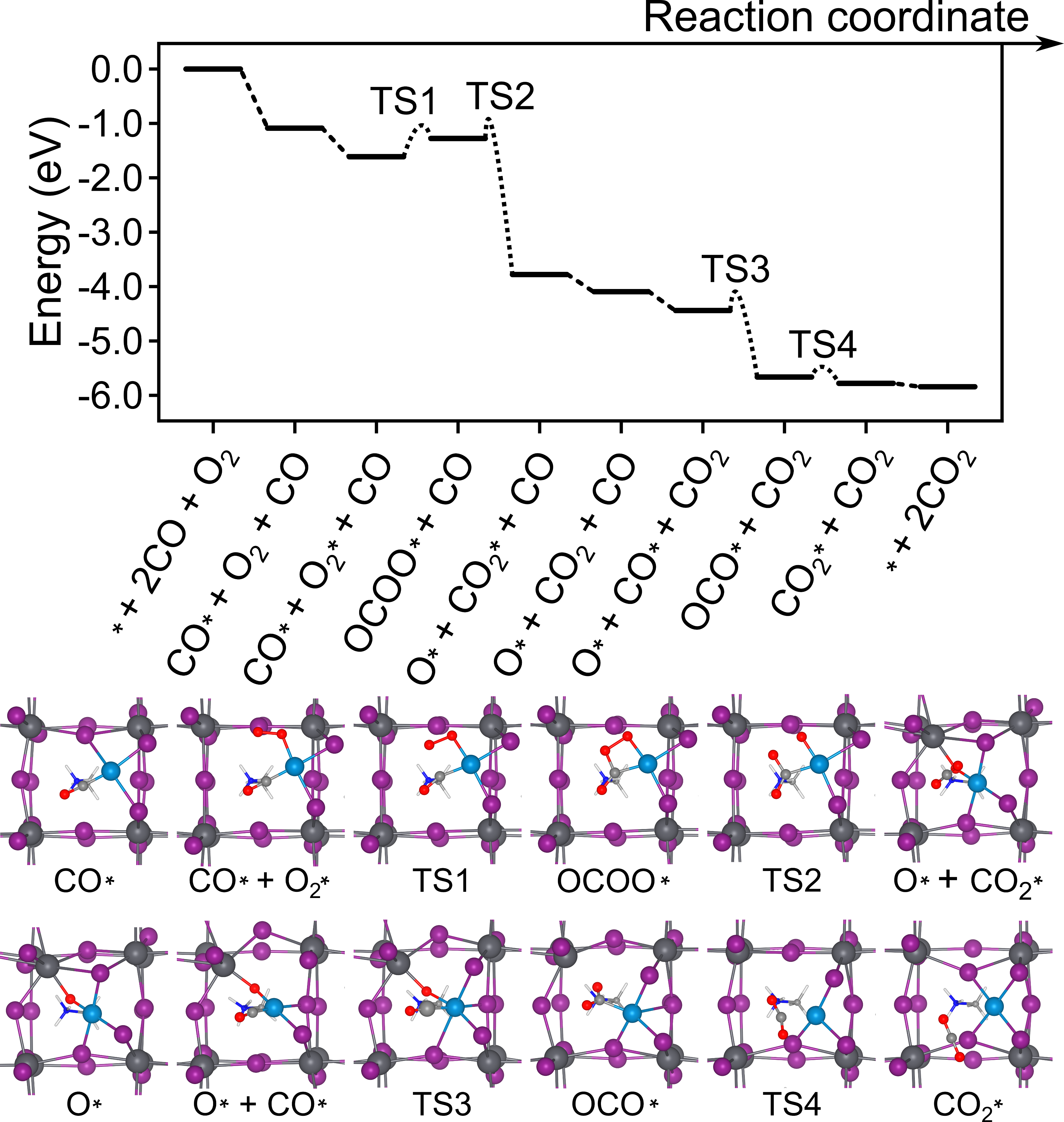}
  \caption{Reaction pathway and free-energy profile of CO oxidation at the Pt$_{1}$@(110) system with temperature and pressure for each gaseous reactant being 298 K and 1 atm, respectively. (top). Configurations of the transition states (TS1-4) and intermediate states (bottom). A star represents an empty surface site or indicates an adsorbed species.}
  \label{PES}
\end{figure}

The reaction pathway and the free-energy profile of the catalytic CO oxidation at Pt$_{1}$@(110) are presented in Fig. \ref{PES}. Here, the standard entropy of small gaseous molecules (\textit{i.e.} O$_{2}$, CO, and CO$_{2}$), estimated to be 2 meV/K \cite{Norskov-book}, is accounted for in their adsorption and desorption. No zero-point energy correction is included in the energy profile, since for the reaction 2CO + O$_{2}$ $\rightarrow$ 2CO$_{2}$, its value of 0.25 eV \cite{table} is negligible compared with the total-energy release of 6.43 eV (DFT energy). From the reaction pathway, one can see that the Pt$_{1}$ site is first occupied by a CO molecule, followed by the adsorption of O$_{2}$ (O$-$O bond length of 1.33 {\AA}), which is consistent with the adsorption energies in Table \ref{Eds}. Here, two Pt-I bonds are broken upon the adsorption of CO and O$_{2}$, leaving the Pt$_{1}$ site bound with only two iodine anions. It is worth noting that breaking the remaining two Pt-I bonds via successive adsorption of CO and O$_{2}$ is energetically highly unfavorable, which indicates that the Pt atom can be stabilized by the MAPbI$_{3}$ host. The O$_{2}$ and CO adsorbates then move toward each other and combine into an OCOO complex after overcoming an energy barrier of 0.58 eV (via TS1). In the complex, the O-O bond is further elongated to 1.51 {\AA} and easily breaks by crossing the second barrier of 0.35 eV (via TS2), generating an O atom and a CO$_{2}$ molecule. After the desorption of CO$_{2}$, another CO molecule adsorbs at the Pt$_{1}$ site, and then, together with the O atom, forms a bent OCO complex by crossing a barrier of 0.39 eV (via TS3). Accompanied by overcoming the fourth barrier of 0.19 eV (via TS4), the OCO complex becomes straight and detaches, leaving the Pt$_{1}$ site available for another round of CO oxidation.

For the MAI-terminated (110) substrate, the Pt single atom could also replace an iodine atom at the outermost layer. The different chemical environment on the surface may affect the deposition and the catalytic properties of the Pt$_{1}$ species. To explore this issue, first-principles calculations were performed where detailed results are presented in the Supplemental Material \cite{SM}. In this case, the adsorption energies of a Pt and an I atom on the host are -4.29 eV and -3.68 eV, respectively, indicating that Pt$_{1}$ can substitute the iodine species on the surface. Upon such substitution, the Pt single atom binds to one Pb and one I atom (left panel of Fig. S7), behaving as an adsorption site for O$_{2}$ and CO molecules. It is interesting that the initial adsorption energies of O$_{2}$ and CO on such Pt atom are -2.73 eV and -2.33 eV, respectively, meaning that the inhibition of O$_{2}$ adsorption by CO no longer exists. Moreover, upon O$_{2}$ adsorption, the molecule carries 0.88 extra electrons and its O-O bond increases from 1.24 {\AA} to 1.43 {\AA}. Thus, the preferential adsorption and subsequent activation of O$_{2}$ molecules still hold. A problem of the Pt single atom by replacing iodine is the too strong interaction with the adsorbates, which may hinder the desorption of reaction products like CO$_{2}$. Such strong interaction comes from a lower coordination number (CN = 2) and a more negative charge (q = -0.63 \textit{e}) of Pt$_{1}$, compared with the corresponding values (CN = 4; q = -0.13 \textit{e}) in the case of replacing MA. In experiments, the substitution of Pt single atoms could be selectively achieved by tuning the reaction conditions that can change the chemical potentials of both iodine atoms and MA groups.

From the bottom right part of the free-energy profile in Fig. \ref{PES}, one can see that the adsorption of CO$_{2}$ and the formation of the OCO complex can take place by crossing a low barrier of 0.31 eV (vis TS4). The bent configuration as well as the negative charge of 0.55 e$^{-}$ carried by the OCO complex demonstrates the activation of CO$_{2}$, which corresponds to the initial step of its reduction \cite{Freund1996}. Reactive hydrogen atoms generated at the neighboring Pt$_{1}$ sites, possibly via feasible H$_{2}$ dissociation \cite{FuQ2013}, are expected to spillover \cite{Conner1995} and hydrogenate the CO$_{2}$ adsorbates, making the Pt$_{1}$@(110) system a potential catalyst toward thermal reduction of CO$_{2}$ \cite{Porosoff2016}. In addition, since MAPbI$_{3}$ is a photoactive material, the electrons that are excited to the conduction band under illumination may transfer to the Pt$_{1}$ sites and trigger processes of CO$_{2}$ photoreduction \cite{Chang2016}. One possible drawback of the material may be related to its long-term sustainability, since HOIPs suffer from decomposition under external stimulus like humidity, light, and heat \cite{Yang2017}. It should be noted, however, that HOIPs have indeed been employed as catalysts for chemical reactions \cite{ParkS2016,XuYF2017,ChenK2017,WuY2018,HuangH2018}, which means that the obstacles to the materials' stability can be overcome. Moreover, significantly enhanced stability of perovskites has been achieved in experiments through compositional engineering by doping different halides or cations \cite{WangZ2017,GongJ2018}, through shape engineering by employing two-dimensional phases \cite{Smith2014,ZhangX2017}, and through interfacial engineering by crystal-crosslinking \cite{Nazeeruddin2015} or forming heterostructures \cite{WangZP2017}. For practical applications in (photo)catalysis, breakthroughs in the stability of HOIPs may be still required, but could be expected considering the tremendous ongoing research efforts.

To conclude, by performing first-principles calculations, we have shown that methylammonium lead iodide could be used as a substrate for single-atom catalysts. By replacing the surface MA groups, isolated Pt single atoms can be implanted and bound by four I$^{-}$ anions, which provide potentially uniform anchoring sites. At the Pt$_{1}$ sites, inhibition of O$_{2}$ adsorption by CO is effectively tackled. Charge transfer toward the O$_{2}$ adsorbate results in its activation that is not sensitive to Pt atomicity or MAPbI$_{3}$ orientation. The supported Pt single atoms are able to catalyze CO oxidation, and may play a role in CO$_{2}$ (photo)reduction reactions. With these findings, we demonstrate the potential application of hybrid perovskites as supporting templates for single-atom catalysts.

\begin{acknowledgments}
We gratefully acknowledge partial support from the European Union's Horizon 2020 research and innovation programme, Grant Agreement No. 676580 through the Center of Excellene NOMAD (Novel Materials Discovery Laboratory, https://NOMAD-CoE.eu) and funding from the Deutsche Forschungsgemeinschaft (DFG) - Projektnummer 182087777 - SFB 951. Q.F. also appreciates support from the Shandong Provincial Natural Science Foundation, China (ZR2018QB005), the Young Scholars Program of Shandong University(2018WLJH49), and the Fundamental Research Funds of Shandong University (2018TB006). We thank Dingsheng Wang, Shufang Ji, and Shubo Tian in Tsinghua University for helpful discussion.
\end{acknowledgments}


\begin{thebibliography}{65}
\expandafter\ifx\csname natexlab\endcsname\relax\def\natexlab#1{#1}\fi
\expandafter\ifx\csname bibnamefont\endcsname\relax
  \def\bibnamefont#1{#1}\fi
\expandafter\ifx\csname bibfnamefont\endcsname\relax
  \def\bibfnamefont#1{#1}\fi
\expandafter\ifx\csname citenamefont\endcsname\relax
  \def\citenamefont#1{#1}\fi
\expandafter\ifx\csname url\endcsname\relax
  \def\url#1{\texttt{#1}}\fi
\expandafter\ifx\csname urlprefix\endcsname\relax\def\urlprefix{URL }\fi
\providecommand{\bibinfo}[2]{#2}
\providecommand{\eprint}[2][]{\url{#2}}

\bibitem[{\citenamefont{Fu et~al.}(2003)\citenamefont{Fu, Saltsburg, and
  Flytzani-Stephanopoulos}}]{Fu2003}
\bibinfo{author}{\bibfnamefont{Q.}~\bibnamefont{Fu}},
  \bibinfo{author}{\bibfnamefont{H.}~\bibnamefont{Saltsburg}},
  \bibnamefont{and}
  \bibinfo{author}{\bibfnamefont{M.}~\bibnamefont{Flytzani-Stephanopoulos}},
  \bibinfo{journal}{Science} \textbf{\bibinfo{volume}{301}},
  \bibinfo{pages}{935} (\bibinfo{year}{2003}).

\bibitem[{\citenamefont{Vajda et~al.}(2009)\citenamefont{Vajda, Pellin,
  Greeley, Marshall, Curtiss, Ballentine, Elam, Catillon-Mucherie, Redfern,
  Mehmood et~al.}}]{Vajda2009}
\bibinfo{author}{\bibfnamefont{S.}~\bibnamefont{Vajda}},
  \bibinfo{author}{\bibfnamefont{M.}~\bibnamefont{Pellin}},
  \bibinfo{author}{\bibfnamefont{J.}~\bibnamefont{Greeley}},
  \bibinfo{author}{\bibfnamefont{C.}~\bibnamefont{Marshall}},
  \bibinfo{author}{\bibfnamefont{L.}~\bibnamefont{Curtiss}},
  \bibinfo{author}{\bibfnamefont{G.}~\bibnamefont{Ballentine}},
  \bibinfo{author}{\bibfnamefont{J.}~\bibnamefont{Elam}},
  \bibinfo{author}{\bibfnamefont{S.}~\bibnamefont{Catillon-Mucherie}},
  \bibinfo{author}{\bibfnamefont{P.}~\bibnamefont{Redfern}},
  \bibinfo{author}{\bibfnamefont{F.}~\bibnamefont{Mehmood}},
  \bibnamefont{et~al.}, \bibinfo{journal}{Nat. Mater.}
  \textbf{\bibinfo{volume}{8}}, \bibinfo{pages}{213} (\bibinfo{year}{2009}).

\bibitem[{\citenamefont{Zhai et~al.}(2010)\citenamefont{Zhai, Pierre, Si, Deng,
  Ferrin, Nilekar, Peng, Herron, Bell, Saltsburg et~al.}}]{ZhaiY2010}
\bibinfo{author}{\bibfnamefont{Y.}~\bibnamefont{Zhai}},
  \bibinfo{author}{\bibfnamefont{D.}~\bibnamefont{Pierre}},
  \bibinfo{author}{\bibfnamefont{R.}~\bibnamefont{Si}},
  \bibinfo{author}{\bibfnamefont{W.}~\bibnamefont{Deng}},
  \bibinfo{author}{\bibfnamefont{P.}~\bibnamefont{Ferrin}},
  \bibinfo{author}{\bibfnamefont{A.~U.} \bibnamefont{Nilekar}},
  \bibinfo{author}{\bibfnamefont{G.}~\bibnamefont{Peng}},
  \bibinfo{author}{\bibfnamefont{J.~A.} \bibnamefont{Herron}},
  \bibinfo{author}{\bibfnamefont{D.~C.} \bibnamefont{Bell}},
  \bibinfo{author}{\bibfnamefont{H.}~\bibnamefont{Saltsburg}},
  \bibnamefont{et~al.}, \bibinfo{journal}{Science}
  \textbf{\bibinfo{volume}{329}}, \bibinfo{pages}{1633} (\bibinfo{year}{2010}).

\bibitem[{\citenamefont{Qiao et~al.}(2011)\citenamefont{Qiao, Wang, Yang,
  Allard, Jiang, Cui, Liu, Li, and Zhang}}]{Qiao2011}
\bibinfo{author}{\bibfnamefont{B.}~\bibnamefont{Qiao}},
  \bibinfo{author}{\bibfnamefont{A.}~\bibnamefont{Wang}},
  \bibinfo{author}{\bibfnamefont{X.}~\bibnamefont{Yang}},
  \bibinfo{author}{\bibfnamefont{L.~F.} \bibnamefont{Allard}},
  \bibinfo{author}{\bibfnamefont{Z.}~\bibnamefont{Jiang}},
  \bibinfo{author}{\bibfnamefont{Y.}~\bibnamefont{Cui}},
  \bibinfo{author}{\bibfnamefont{J.}~\bibnamefont{Liu}},
  \bibinfo{author}{\bibfnamefont{J.}~\bibnamefont{Li}}, \bibnamefont{and}
  \bibinfo{author}{\bibfnamefont{T.}~\bibnamefont{Zhang}},
  \bibinfo{journal}{Nat. Chem.} \textbf{\bibinfo{volume}{3}},
  \bibinfo{pages}{634} (\bibinfo{year}{2011}).

\bibitem[{\citenamefont{Moses-DeBusk et~al.}(2013)\citenamefont{Moses-DeBusk,
  Yoon, Allard, Mullins, Wu, Yang, Veith, Stocks, and
  Narula}}]{Moses-DeBusk2013}
\bibinfo{author}{\bibfnamefont{M.}~\bibnamefont{Moses-DeBusk}},
  \bibinfo{author}{\bibfnamefont{M.}~\bibnamefont{Yoon}},
  \bibinfo{author}{\bibfnamefont{L.~F.} \bibnamefont{Allard}},
  \bibinfo{author}{\bibfnamefont{D.~R.} \bibnamefont{Mullins}},
  \bibinfo{author}{\bibfnamefont{Z.}~\bibnamefont{Wu}},
  \bibinfo{author}{\bibfnamefont{X.}~\bibnamefont{Yang}},
  \bibinfo{author}{\bibfnamefont{G.}~\bibnamefont{Veith}},
  \bibinfo{author}{\bibfnamefont{G.~M.} \bibnamefont{Stocks}},
  \bibnamefont{and} \bibinfo{author}{\bibfnamefont{C.~K.}
  \bibnamefont{Narula}}, \bibinfo{journal}{J. Am. Chem. Soc.}
  \textbf{\bibinfo{volume}{135}}, \bibinfo{pages}{12634}
  (\bibinfo{year}{2013}).

\bibitem[{\citenamefont{Kistler et~al.}(2014)\citenamefont{Kistler, Chotigkrai,
  Xu, Enderle, Praserthdam, Chen, Browning, and Gates}}]{Kistler2014}
\bibinfo{author}{\bibfnamefont{J.~D.} \bibnamefont{Kistler}},
  \bibinfo{author}{\bibfnamefont{N.}~\bibnamefont{Chotigkrai}},
  \bibinfo{author}{\bibfnamefont{P.}~\bibnamefont{Xu}},
  \bibinfo{author}{\bibfnamefont{B.}~\bibnamefont{Enderle}},
  \bibinfo{author}{\bibfnamefont{P.}~\bibnamefont{Praserthdam}},
  \bibinfo{author}{\bibfnamefont{C.-Y.} \bibnamefont{Chen}},
  \bibinfo{author}{\bibfnamefont{N.~D.} \bibnamefont{Browning}},
  \bibnamefont{and} \bibinfo{author}{\bibfnamefont{B.~C.} \bibnamefont{Gates}},
  \bibinfo{journal}{Angew. Chem. Int. Ed.} \textbf{\bibinfo{volume}{53}},
  \bibinfo{pages}{8904} (\bibinfo{year}{2014}).

\bibitem[{\citenamefont{Ding et~al.}(2015)\citenamefont{Ding, Gulec, Johnson,
  Schweitzer, Stucky, Marks, and Stair}}]{DingK2015}
\bibinfo{author}{\bibfnamefont{K.}~\bibnamefont{Ding}},
  \bibinfo{author}{\bibfnamefont{A.}~\bibnamefont{Gulec}},
  \bibinfo{author}{\bibfnamefont{A.~M.} \bibnamefont{Johnson}},
  \bibinfo{author}{\bibfnamefont{N.~M.} \bibnamefont{Schweitzer}},
  \bibinfo{author}{\bibfnamefont{G.~D.} \bibnamefont{Stucky}},
  \bibinfo{author}{\bibfnamefont{L.~D.} \bibnamefont{Marks}}, \bibnamefont{and}
  \bibinfo{author}{\bibfnamefont{P.~C.} \bibnamefont{Stair}},
  \bibinfo{journal}{Science} \textbf{\bibinfo{volume}{350}},
  \bibinfo{pages}{189} (\bibinfo{year}{2015}).

\bibitem[{\citenamefont{Dvo{\v{r}}{\'{a}}k
  et~al.}(2016)\citenamefont{Dvo{\v{r}}{\'{a}}k, {Farnesi Camellone}, Tovt,
  Tran, Negreiros, Vorokhta, Sk{\'{a}}la, Matol{\'{i}}nov{\'{a}},
  Myslive{\v{c}}ek, Matol{\'{i}}n et~al.}}]{Dvorak2016}
\bibinfo{author}{\bibfnamefont{F.}~\bibnamefont{Dvo{\v{r}}{\'{a}}k}},
  \bibinfo{author}{\bibfnamefont{M.}~\bibnamefont{{Farnesi Camellone}}},
  \bibinfo{author}{\bibfnamefont{A.}~\bibnamefont{Tovt}},
  \bibinfo{author}{\bibfnamefont{N.-D.} \bibnamefont{Tran}},
  \bibinfo{author}{\bibfnamefont{F.~R.} \bibnamefont{Negreiros}},
  \bibinfo{author}{\bibfnamefont{M.}~\bibnamefont{Vorokhta}},
  \bibinfo{author}{\bibfnamefont{T.}~\bibnamefont{Sk{\'{a}}la}},
  \bibinfo{author}{\bibfnamefont{I.}~\bibnamefont{Matol{\'{i}}nov{\'{a}}}},
  \bibinfo{author}{\bibfnamefont{J.}~\bibnamefont{Myslive{\v{c}}ek}},
  \bibinfo{author}{\bibfnamefont{V.}~\bibnamefont{Matol{\'{i}}n}},
  \bibnamefont{et~al.}, \bibinfo{journal}{Nat. Commun.}
  \textbf{\bibinfo{volume}{7}}, \bibinfo{pages}{10801} (\bibinfo{year}{2016}).

\bibitem[{\citenamefont{Gao et~al.}(2016)\citenamefont{Gao, Jiao, Waclawik, and
  Du}}]{Gao2016}
\bibinfo{author}{\bibfnamefont{G.}~\bibnamefont{Gao}},
  \bibinfo{author}{\bibfnamefont{Y.}~\bibnamefont{Jiao}},
  \bibinfo{author}{\bibfnamefont{E.~R.} \bibnamefont{Waclawik}},
  \bibnamefont{and} \bibinfo{author}{\bibfnamefont{A.}~\bibnamefont{Du}},
  \bibinfo{journal}{J. Am. Chem. Soc.} \textbf{\bibinfo{volume}{138}},
  \bibinfo{pages}{6292} (\bibinfo{year}{2016}).

\bibitem[{\citenamefont{Nie et~al.}(2017)\citenamefont{Nie, Mei, Xiong, Peng,
  Ren, Hernandez, DeLaRiva, Wang, Engelhard, Kovarik et~al.}}]{Nie2017}
\bibinfo{author}{\bibfnamefont{L.}~\bibnamefont{Nie}},
  \bibinfo{author}{\bibfnamefont{D.}~\bibnamefont{Mei}},
  \bibinfo{author}{\bibfnamefont{H.}~\bibnamefont{Xiong}},
  \bibinfo{author}{\bibfnamefont{B.}~\bibnamefont{Peng}},
  \bibinfo{author}{\bibfnamefont{Z.}~\bibnamefont{Ren}},
  \bibinfo{author}{\bibfnamefont{X.~I.~P.} \bibnamefont{Hernandez}},
  \bibinfo{author}{\bibfnamefont{A.}~\bibnamefont{DeLaRiva}},
  \bibinfo{author}{\bibfnamefont{M.}~\bibnamefont{Wang}},
  \bibinfo{author}{\bibfnamefont{M.~H.} \bibnamefont{Engelhard}},
  \bibinfo{author}{\bibfnamefont{L.}~\bibnamefont{Kovarik}},
  \bibnamefont{et~al.}, \bibinfo{journal}{Science}
  \textbf{\bibinfo{volume}{358}}, \bibinfo{pages}{1419} (\bibinfo{year}{2017}).

\bibitem[{\citenamefont{Wei et~al.}(2018)\citenamefont{Wei, Li, Liu, Li, Chen,
  Gong, Zhang, Cheong, Wang, Zheng et~al.}}]{Wei2018}
\bibinfo{author}{\bibfnamefont{S.}~\bibnamefont{Wei}},
  \bibinfo{author}{\bibfnamefont{A.}~\bibnamefont{Li}},
  \bibinfo{author}{\bibfnamefont{J.-C.} \bibnamefont{Liu}},
  \bibinfo{author}{\bibfnamefont{Z.}~\bibnamefont{Li}},
  \bibinfo{author}{\bibfnamefont{W.}~\bibnamefont{Chen}},
  \bibinfo{author}{\bibfnamefont{Y.}~\bibnamefont{Gong}},
  \bibinfo{author}{\bibfnamefont{Q.}~\bibnamefont{Zhang}},
  \bibinfo{author}{\bibfnamefont{W.-C.} \bibnamefont{Cheong}},
  \bibinfo{author}{\bibfnamefont{Y.}~\bibnamefont{Wang}},
  \bibinfo{author}{\bibfnamefont{L.}~\bibnamefont{Zheng}},
  \bibnamefont{et~al.}, \bibinfo{journal}{Nat. Nanotechnol.}
  \textbf{\bibinfo{volume}{13}}, \bibinfo{pages}{856} (\bibinfo{year}{2018}).

\bibitem[{\citenamefont{Therrien et~al.}(2018)\citenamefont{Therrien, Hensley,
  Marcinkowski, Zhang, Lucci, Coughlin, Schilling, McEwen, and
  Sykes}}]{Therrien2018}
\bibinfo{author}{\bibfnamefont{A.~J.} \bibnamefont{Therrien}},
  \bibinfo{author}{\bibfnamefont{A.~J.~R.} \bibnamefont{Hensley}},
  \bibinfo{author}{\bibfnamefont{M.~D.} \bibnamefont{Marcinkowski}},
  \bibinfo{author}{\bibfnamefont{R.}~\bibnamefont{Zhang}},
  \bibinfo{author}{\bibfnamefont{F.~R.} \bibnamefont{Lucci}},
  \bibinfo{author}{\bibfnamefont{B.}~\bibnamefont{Coughlin}},
  \bibinfo{author}{\bibfnamefont{A.~C.} \bibnamefont{Schilling}},
  \bibinfo{author}{\bibfnamefont{J.-S.} \bibnamefont{McEwen}},
  \bibnamefont{and} \bibinfo{author}{\bibfnamefont{E.~C.~H.}
  \bibnamefont{Sykes}}, \bibinfo{journal}{Nat. Catal.}
  \textbf{\bibinfo{volume}{1}}, \bibinfo{pages}{192} (\bibinfo{year}{2018}).

\bibitem[{\citenamefont{Liu}(2017)}]{LiuJY2017}
\bibinfo{author}{\bibfnamefont{J.}~\bibnamefont{Liu}}, \bibinfo{journal}{ACS
  Catal.} \textbf{\bibinfo{volume}{7}}, \bibinfo{pages}{34}
  (\bibinfo{year}{2017}).

\bibitem[{\citenamefont{Wang et~al.}(2018)\citenamefont{Wang, Li, and
  Zhang}}]{Wang2018}
\bibinfo{author}{\bibfnamefont{A.}~\bibnamefont{Wang}},
  \bibinfo{author}{\bibfnamefont{J.}~\bibnamefont{Li}}, \bibnamefont{and}
  \bibinfo{author}{\bibfnamefont{T.}~\bibnamefont{Zhang}},
  \bibinfo{journal}{Nat. Rev. Chem.} \textbf{\bibinfo{volume}{2}},
  \bibinfo{pages}{65} (\bibinfo{year}{2018}).

\bibitem[{\citenamefont{Yang and Gates}(2017)}]{YangD2017}
\bibinfo{author}{\bibfnamefont{D.}~\bibnamefont{Yang}} \bibnamefont{and}
  \bibinfo{author}{\bibfnamefont{B.~C.} \bibnamefont{Gates}},
  \bibinfo{journal}{Nat. Mater.} \textbf{\bibinfo{volume}{16}},
  \bibinfo{pages}{703} (\bibinfo{year}{2017}).

\bibitem[{\citenamefont{Fang et~al.}(2018)\citenamefont{Fang, Shang, Wang,
  Jiao, Yao, Li, Zhang, Luo, and Jiang}}]{Fang2018}
\bibinfo{author}{\bibfnamefont{X.}~\bibnamefont{Fang}},
  \bibinfo{author}{\bibfnamefont{Q.}~\bibnamefont{Shang}},
  \bibinfo{author}{\bibfnamefont{Y.}~\bibnamefont{Wang}},
  \bibinfo{author}{\bibfnamefont{L.}~\bibnamefont{Jiao}},
  \bibinfo{author}{\bibfnamefont{T.}~\bibnamefont{Yao}},
  \bibinfo{author}{\bibfnamefont{Y.}~\bibnamefont{Li}},
  \bibinfo{author}{\bibfnamefont{Q.}~\bibnamefont{Zhang}},
  \bibinfo{author}{\bibfnamefont{Y.}~\bibnamefont{Luo}}, \bibnamefont{and}
  \bibinfo{author}{\bibfnamefont{H.-L.} \bibnamefont{Jiang}},
  \bibinfo{journal}{Adv. Mater.} \textbf{\bibinfo{volume}{30}},
  \bibinfo{pages}{1705112} (\bibinfo{year}{2018}).

\bibitem[{\citenamefont{Walsh}(2015)}]{Walsh2015}
\bibinfo{author}{\bibfnamefont{A.}~\bibnamefont{Walsh}}, \bibinfo{journal}{J.
  Phys. Chem. C} \textbf{\bibinfo{volume}{119}}, \bibinfo{pages}{5755}
  (\bibinfo{year}{2015}).

\bibitem[{\citenamefont{Yin et~al.}(2015)\citenamefont{Yin, Yang, Kang, Yan,
  and Wei}}]{Yin2014}
\bibinfo{author}{\bibfnamefont{W.-J.} \bibnamefont{Yin}},
  \bibinfo{author}{\bibfnamefont{J.-H.} \bibnamefont{Yang}},
  \bibinfo{author}{\bibfnamefont{J.}~\bibnamefont{Kang}},
  \bibinfo{author}{\bibfnamefont{Y.}~\bibnamefont{Yan}}, \bibnamefont{and}
  \bibinfo{author}{\bibfnamefont{S.-H.} \bibnamefont{Wei}},
  \bibinfo{journal}{J. Mater. Chem. A} \textbf{\bibinfo{volume}{3}},
  \bibinfo{pages}{8926} (\bibinfo{year}{2015}).

\bibitem[{\citenamefont{Colella et~al.}(2016)\citenamefont{Colella, Mazzeo,
  Rizzo, Gigli, and Listorti}}]{Colella2016}
\bibinfo{author}{\bibfnamefont{S.}~\bibnamefont{Colella}},
  \bibinfo{author}{\bibfnamefont{M.}~\bibnamefont{Mazzeo}},
  \bibinfo{author}{\bibfnamefont{A.}~\bibnamefont{Rizzo}},
  \bibinfo{author}{\bibfnamefont{G.}~\bibnamefont{Gigli}}, \bibnamefont{and}
  \bibinfo{author}{\bibfnamefont{A.}~\bibnamefont{Listorti}},
  \bibinfo{journal}{J. Phys. Chem. Lett.} \textbf{\bibinfo{volume}{7}},
  \bibinfo{pages}{4322} (\bibinfo{year}{2016}).

\bibitem[{\citenamefont{Park et~al.}(2016)\citenamefont{Park, Chang, Lee, Park,
  Ahn, and Nam}}]{ParkS2016}
\bibinfo{author}{\bibfnamefont{S.}~\bibnamefont{Park}},
  \bibinfo{author}{\bibfnamefont{W.~J.} \bibnamefont{Chang}},
  \bibinfo{author}{\bibfnamefont{C.~W.} \bibnamefont{Lee}},
  \bibinfo{author}{\bibfnamefont{S.}~\bibnamefont{Park}},
  \bibinfo{author}{\bibfnamefont{H.-Y.} \bibnamefont{Ahn}}, \bibnamefont{and}
  \bibinfo{author}{\bibfnamefont{K.~T.} \bibnamefont{Nam}},
  \bibinfo{journal}{Nat. Energy} \textbf{\bibinfo{volume}{2}},
  \bibinfo{pages}{16185} (\bibinfo{year}{2016}).

\bibitem[{\citenamefont{Xu et~al.}(2017)\citenamefont{Xu, Yang, Chen, Wang,
  Chen, Kuang, and Su}}]{XuYF2017}
\bibinfo{author}{\bibfnamefont{Y.-F.} \bibnamefont{Xu}},
  \bibinfo{author}{\bibfnamefont{M.-Z.} \bibnamefont{Yang}},
  \bibinfo{author}{\bibfnamefont{B.-X.} \bibnamefont{Chen}},
  \bibinfo{author}{\bibfnamefont{X.-D.} \bibnamefont{Wang}},
  \bibinfo{author}{\bibfnamefont{H.-Y.} \bibnamefont{Chen}},
  \bibinfo{author}{\bibfnamefont{D.-B.} \bibnamefont{Kuang}}, \bibnamefont{and}
  \bibinfo{author}{\bibfnamefont{C.-Y.} \bibnamefont{Su}}, \bibinfo{journal}{J.
  Am. Chem. Soc.} \textbf{\bibinfo{volume}{139}}, \bibinfo{pages}{5660}
  (\bibinfo{year}{2017}).

\bibitem[{\citenamefont{Chen et~al.}(2017)\citenamefont{Chen, Deng, Dodekatos,
  and T{\"{u}}ys{\"{u}}z}}]{ChenK2017}
\bibinfo{author}{\bibfnamefont{K.}~\bibnamefont{Chen}},
  \bibinfo{author}{\bibfnamefont{X.}~\bibnamefont{Deng}},
  \bibinfo{author}{\bibfnamefont{G.}~\bibnamefont{Dodekatos}},
  \bibnamefont{and}
  \bibinfo{author}{\bibfnamefont{H.}~\bibnamefont{T{\"{u}}ys{\"{u}}z}},
  \bibinfo{journal}{J. Am. Chem. Soc.} \textbf{\bibinfo{volume}{139}},
  \bibinfo{pages}{12267} (\bibinfo{year}{2017}).

\bibitem[{\citenamefont{Wu et~al.}(2018)\citenamefont{Wu, Wang, Zhu, Zhang,
  Wang, Liu, Zou, Dai, Whangbo, and Huang}}]{WuY2018}
\bibinfo{author}{\bibfnamefont{Y.}~\bibnamefont{Wu}},
  \bibinfo{author}{\bibfnamefont{P.}~\bibnamefont{Wang}},
  \bibinfo{author}{\bibfnamefont{X.}~\bibnamefont{Zhu}},
  \bibinfo{author}{\bibfnamefont{Q.}~\bibnamefont{Zhang}},
  \bibinfo{author}{\bibfnamefont{Z.}~\bibnamefont{Wang}},
  \bibinfo{author}{\bibfnamefont{Y.}~\bibnamefont{Liu}},
  \bibinfo{author}{\bibfnamefont{G.}~\bibnamefont{Zou}},
  \bibinfo{author}{\bibfnamefont{Y.}~\bibnamefont{Dai}},
  \bibinfo{author}{\bibfnamefont{M.~H.} \bibnamefont{Whangbo}},
  \bibnamefont{and} \bibinfo{author}{\bibfnamefont{B.}~\bibnamefont{Huang}},
  \bibinfo{journal}{Adv. Mater.} \textbf{\bibinfo{volume}{30}},
  \bibinfo{pages}{2} (\bibinfo{year}{2018}).

\bibitem[{\citenamefont{Huang et~al.}(2018)\citenamefont{Huang, Yuan, Janssen,
  Sol{\'{i}}s-Fern{\'{a}}ndez, Wang, Tan, Jonckheere, Debroye, Long, Hendrix
  et~al.}}]{HuangH2018}
\bibinfo{author}{\bibfnamefont{H.}~\bibnamefont{Huang}},
  \bibinfo{author}{\bibfnamefont{H.}~\bibnamefont{Yuan}},
  \bibinfo{author}{\bibfnamefont{K.~P.} \bibnamefont{Janssen}},
  \bibinfo{author}{\bibfnamefont{G.}~\bibnamefont{Sol{\'{i}}s-Fern{\'{a}}ndez}},
  \bibinfo{author}{\bibfnamefont{Y.}~\bibnamefont{Wang}},
  \bibinfo{author}{\bibfnamefont{C.~Y.} \bibnamefont{Tan}},
  \bibinfo{author}{\bibfnamefont{D.}~\bibnamefont{Jonckheere}},
  \bibinfo{author}{\bibfnamefont{E.}~\bibnamefont{Debroye}},
  \bibinfo{author}{\bibfnamefont{J.}~\bibnamefont{Long}},
  \bibinfo{author}{\bibfnamefont{J.}~\bibnamefont{Hendrix}},
  \bibnamefont{et~al.}, \bibinfo{journal}{ACS Energy Lett.}
  \textbf{\bibinfo{volume}{3}}, \bibinfo{pages}{755} (\bibinfo{year}{2018}).

\bibitem[{\citenamefont{Ertl}(1994)}]{Ertl1994}
\bibinfo{author}{\bibfnamefont{G.}~\bibnamefont{Ertl}}, \bibinfo{journal}{Surf.
  Sci.} \textbf{\bibinfo{volume}{299-300}}, \bibinfo{pages}{742}
  (\bibinfo{year}{1994}).

\bibitem[{\citenamefont{Freund et~al.}(2011)\citenamefont{Freund, Meijer,
  Scheffler, Schl\"{o}gl, and Wolf}}]{Freund2011}
\bibinfo{author}{\bibfnamefont{H.-J.} \bibnamefont{Freund}},
  \bibinfo{author}{\bibfnamefont{G.}~\bibnamefont{Meijer}},
  \bibinfo{author}{\bibfnamefont{M.}~\bibnamefont{Scheffler}},
  \bibinfo{author}{\bibfnamefont{R.}~\bibnamefont{Schl\"{o}gl}},
  \bibnamefont{and} \bibinfo{author}{\bibfnamefont{M.}~\bibnamefont{Wolf}},
  \bibinfo{journal}{Angew. Chem. Int. Ed.} \textbf{\bibinfo{volume}{50}},
  \bibinfo{pages}{10064} (\bibinfo{year}{2011}).

\bibitem[{\citenamefont{Newton et~al.}(2015)\citenamefont{Newton, Ferri,
  Smolentsev, Marchionni, and Nachtegaal}}]{Newton2015}
\bibinfo{author}{\bibfnamefont{M.~A.} \bibnamefont{Newton}},
  \bibinfo{author}{\bibfnamefont{D.}~\bibnamefont{Ferri}},
  \bibinfo{author}{\bibfnamefont{G.}~\bibnamefont{Smolentsev}},
  \bibinfo{author}{\bibfnamefont{V.}~\bibnamefont{Marchionni}},
  \bibnamefont{and}
  \bibinfo{author}{\bibfnamefont{M.}~\bibnamefont{Nachtegaal}},
  \bibinfo{journal}{Nat. Commun.} \textbf{\bibinfo{volume}{6}},
  \bibinfo{pages}{8675} (\bibinfo{year}{2015}).

\bibitem[{\citenamefont{Gatla et~al.}(2016)\citenamefont{Gatla, Aubert,
  Agostini, Mathon, Pascarelli, Lunkenbein, Willinger, and Kaper}}]{Gatla2016}
\bibinfo{author}{\bibfnamefont{S.}~\bibnamefont{Gatla}},
  \bibinfo{author}{\bibfnamefont{D.}~\bibnamefont{Aubert}},
  \bibinfo{author}{\bibfnamefont{G.}~\bibnamefont{Agostini}},
  \bibinfo{author}{\bibfnamefont{O.}~\bibnamefont{Mathon}},
  \bibinfo{author}{\bibfnamefont{S.}~\bibnamefont{Pascarelli}},
  \bibinfo{author}{\bibfnamefont{T.}~\bibnamefont{Lunkenbein}},
  \bibinfo{author}{\bibfnamefont{M.~G.} \bibnamefont{Willinger}},
  \bibnamefont{and} \bibinfo{author}{\bibfnamefont{H.}~\bibnamefont{Kaper}},
  \bibinfo{journal}{ACS Catal.} \textbf{\bibinfo{volume}{6}},
  \bibinfo{pages}{6151} (\bibinfo{year}{2016}).

\bibitem[{\citenamefont{Christians et~al.}(2015)\citenamefont{Christians,
  Miranda~Herrera, and Kamat}}]{Christians2015}
\bibinfo{author}{\bibfnamefont{J.~A.} \bibnamefont{Christians}},
  \bibinfo{author}{\bibfnamefont{P.~A.} \bibnamefont{Miranda~Herrera}},
  \bibnamefont{and} \bibinfo{author}{\bibfnamefont{P.~V.} \bibnamefont{Kamat}},
  \bibinfo{journal}{J. Am. Chem. Soc.} \textbf{\bibinfo{volume}{137}},
  \bibinfo{pages}{1530} (\bibinfo{year}{2015}).

\bibitem[{\citenamefont{Mosconi et~al.}(2015)\citenamefont{Mosconi, Azpiroz,
  and De~Angelis}}]{Mosconi2015}
\bibinfo{author}{\bibfnamefont{E.}~\bibnamefont{Mosconi}},
  \bibinfo{author}{\bibfnamefont{J.~M.} \bibnamefont{Azpiroz}},
  \bibnamefont{and}
  \bibinfo{author}{\bibfnamefont{F.}~\bibnamefont{De~Angelis}},
  \bibinfo{journal}{Chem. Mater.} \textbf{\bibinfo{volume}{27}},
  \bibinfo{pages}{4885} (\bibinfo{year}{2015}).

\bibitem[{\citenamefont{Zhang et~al.}(2016)\citenamefont{Zhang, Ju, and
  Liang}}]{Zhang2016}
\bibinfo{author}{\bibfnamefont{L.}~\bibnamefont{Zhang}},
  \bibinfo{author}{\bibfnamefont{M.-G.} \bibnamefont{Ju}}, \bibnamefont{and}
  \bibinfo{author}{\bibfnamefont{W.}~\bibnamefont{Liang}},
  \bibinfo{journal}{Phys. Chem. Chem. Phys.} \textbf{\bibinfo{volume}{18}},
  \bibinfo{pages}{23174} (\bibinfo{year}{2016}).

\bibitem[{\citenamefont{Yang and Kelly}(2017)}]{Yang2017}
\bibinfo{author}{\bibfnamefont{J.}~\bibnamefont{Yang}} \bibnamefont{and}
  \bibinfo{author}{\bibfnamefont{T.~L.} \bibnamefont{Kelly}},
  \bibinfo{journal}{Inorg. Chem.} \textbf{\bibinfo{volume}{56}},
  \bibinfo{pages}{92} (\bibinfo{year}{2017}).

\bibitem[{\citenamefont{Quarti et~al.}(2017)\citenamefont{Quarti, {De Angelis},
  and Beljonne}}]{Quarti2017}
\bibinfo{author}{\bibfnamefont{C.}~\bibnamefont{Quarti}},
  \bibinfo{author}{\bibfnamefont{F.}~\bibnamefont{{De Angelis}}},
  \bibnamefont{and} \bibinfo{author}{\bibfnamefont{D.}~\bibnamefont{Beljonne}},
  \bibinfo{journal}{Chem. Mater.} \textbf{\bibinfo{volume}{29}},
  \bibinfo{pages}{958} (\bibinfo{year}{2017}).

\bibitem[{\citenamefont{Geng et~al.}(2015)\citenamefont{Geng, Tong, Tang, Yam,
  Zhang, Lau, and Liu}}]{GengW2015}
\bibinfo{author}{\bibfnamefont{W.}~\bibnamefont{Geng}},
  \bibinfo{author}{\bibfnamefont{C.-J.} \bibnamefont{Tong}},
  \bibinfo{author}{\bibfnamefont{Z.-K.} \bibnamefont{Tang}},
  \bibinfo{author}{\bibfnamefont{C.}~\bibnamefont{Yam}},
  \bibinfo{author}{\bibfnamefont{Y.-N.} \bibnamefont{Zhang}},
  \bibinfo{author}{\bibfnamefont{W.-M.} \bibnamefont{Lau}}, \bibnamefont{and}
  \bibinfo{author}{\bibfnamefont{L.-M.} \bibnamefont{Liu}},
  \bibinfo{journal}{J. Materiomics} \textbf{\bibinfo{volume}{1}},
  \bibinfo{pages}{213} (\bibinfo{year}{2015}).

\bibitem[{\citenamefont{She et~al.}(2016)\citenamefont{She, Liu, and
  Zhong}}]{SheL2016}
\bibinfo{author}{\bibfnamefont{L.}~\bibnamefont{She}},
  \bibinfo{author}{\bibfnamefont{M.}~\bibnamefont{Liu}}, \bibnamefont{and}
  \bibinfo{author}{\bibfnamefont{D.}~\bibnamefont{Zhong}},
  \bibinfo{journal}{ACS Nano} \textbf{\bibinfo{volume}{10}},
  \bibinfo{pages}{1126} (\bibinfo{year}{2016}).

\bibitem[{\citenamefont{Ohmann et~al.}(2015)\citenamefont{Ohmann, Ono, Kim,
  Lin, Lee, Li, Park, and Qi}}]{Ohmann2015}
\bibinfo{author}{\bibfnamefont{R.}~\bibnamefont{Ohmann}},
  \bibinfo{author}{\bibfnamefont{L.~K.} \bibnamefont{Ono}},
  \bibinfo{author}{\bibfnamefont{H.-S.} \bibnamefont{Kim}},
  \bibinfo{author}{\bibfnamefont{H.}~\bibnamefont{Lin}},
  \bibinfo{author}{\bibfnamefont{M.~V.} \bibnamefont{Lee}},
  \bibinfo{author}{\bibfnamefont{Y.}~\bibnamefont{Li}},
  \bibinfo{author}{\bibfnamefont{N.-G.} \bibnamefont{Park}}, \bibnamefont{and}
  \bibinfo{author}{\bibfnamefont{Y.}~\bibnamefont{Qi}}, \bibinfo{journal}{J.
  Am. Chem. Soc.} \textbf{\bibinfo{volume}{137}}, \bibinfo{pages}{16049}
  (\bibinfo{year}{2015}).

\bibitem[{\citenamefont{Kawamura et~al.}(2002)\citenamefont{Kawamura,
  Mashiyama, and Hasebe}}]{Kawamura2002}
\bibinfo{author}{\bibfnamefont{Y.}~\bibnamefont{Kawamura}},
  \bibinfo{author}{\bibfnamefont{H.}~\bibnamefont{Mashiyama}},
  \bibnamefont{and} \bibinfo{author}{\bibfnamefont{K.}~\bibnamefont{Hasebe}},
  \bibinfo{journal}{J. Phys. Soc. Japan} \textbf{\bibinfo{volume}{71}},
  \bibinfo{pages}{1694} (\bibinfo{year}{2002}).

\bibitem[{\citenamefont{Kresse and
  Furthm\"{u}ller}(1996{\natexlab{a}})}]{vasp1}
\bibinfo{author}{\bibfnamefont{G.}~\bibnamefont{Kresse}} \bibnamefont{and}
  \bibinfo{author}{\bibfnamefont{J.}~\bibnamefont{Furthm\"{u}ller}},
  \bibinfo{journal}{{J}. {C}omput. {M}ater. {S}ci.}
  \textbf{\bibinfo{volume}{6}}, \bibinfo{pages}{15}
  (\bibinfo{year}{1996}{\natexlab{a}}).

\bibitem[{\citenamefont{Kresse and
  Furthm\"{u}ller}(1996{\natexlab{b}})}]{vasp2}
\bibinfo{author}{\bibfnamefont{G.}~\bibnamefont{Kresse}} \bibnamefont{and}
  \bibinfo{author}{\bibfnamefont{J.}~\bibnamefont{Furthm\"{u}ller}},
  \bibinfo{journal}{{P}hys. {R}ev. {B}} \textbf{\bibinfo{volume}{54}},
  \bibinfo{pages}{11169} (\bibinfo{year}{1996}{\natexlab{b}}).

\bibitem[{\citenamefont{Bl\"{o}chl}(1994)}]{PAW}
\bibinfo{author}{\bibfnamefont{P.~E.} \bibnamefont{Bl\"{o}chl}},
  \bibinfo{journal}{{P}hys. {R}ev. {B}} \textbf{\bibinfo{volume}{50}},
  \bibinfo{pages}{17953} (\bibinfo{year}{1994}).

\bibitem[{\citenamefont{Dion et~al.}(2004)\citenamefont{Dion, Rydberg,
  Schr\"oder, Langreth, and Lundqvist}}]{Dion04}
\bibinfo{author}{\bibfnamefont{M.}~\bibnamefont{Dion}},
  \bibinfo{author}{\bibfnamefont{H.}~\bibnamefont{Rydberg}},
  \bibinfo{author}{\bibfnamefont{E.}~\bibnamefont{Schr\"oder}},
  \bibinfo{author}{\bibfnamefont{D.~C.} \bibnamefont{Langreth}},
  \bibnamefont{and} \bibinfo{author}{\bibfnamefont{B.~I.}
  \bibnamefont{Lundqvist}}, \bibinfo{journal}{Phys. Rev. Lett.}
  \textbf{\bibinfo{volume}{92}}, \bibinfo{pages}{246401}
  (\bibinfo{year}{2004}).

\bibitem[{\citenamefont{Klime\v{s} et~al.}(2011)\citenamefont{Klime\v{s},
  Bowler, and Michaelides}}]{Klimes11}
\bibinfo{author}{\bibfnamefont{J.}~\bibnamefont{Klime\v{s}}},
  \bibinfo{author}{\bibfnamefont{D.~R.} \bibnamefont{Bowler}},
  \bibnamefont{and}
  \bibinfo{author}{\bibfnamefont{A.}~\bibnamefont{Michaelides}},
  \bibinfo{journal}{Phys. Rev. B} \textbf{\bibinfo{volume}{83}},
  \bibinfo{pages}{195131} (\bibinfo{year}{2011}).

\bibitem[{\citenamefont{Wang et~al.}(2014)\citenamefont{Wang, Gould, Dobson,
  Zhang, Yang, Yao, and Zhao}}]{WangY2014}
\bibinfo{author}{\bibfnamefont{Y.}~\bibnamefont{Wang}},
  \bibinfo{author}{\bibfnamefont{T.}~\bibnamefont{Gould}},
  \bibinfo{author}{\bibfnamefont{J.~F.} \bibnamefont{Dobson}},
  \bibinfo{author}{\bibfnamefont{H.}~\bibnamefont{Zhang}},
  \bibinfo{author}{\bibfnamefont{H.}~\bibnamefont{Yang}},
  \bibinfo{author}{\bibfnamefont{X.}~\bibnamefont{Yao}}, \bibnamefont{and}
  \bibinfo{author}{\bibfnamefont{H.}~\bibnamefont{Zhao}},
  \bibinfo{journal}{Phys. Chem. Chem. Phys.} \textbf{\bibinfo{volume}{16}},
  \bibinfo{pages}{1424} (\bibinfo{year}{2014}).

\bibitem[{\citenamefont{Monkhorst and Pack}(1976)}]{MP}
\bibinfo{author}{\bibfnamefont{H.~J.} \bibnamefont{Monkhorst}}
  \bibnamefont{and} \bibinfo{author}{\bibfnamefont{J.~D.} \bibnamefont{Pack}},
  \bibinfo{journal}{Phys. Rev. B} \textbf{\bibinfo{volume}{13}},
  \bibinfo{pages}{5188} (\bibinfo{year}{1976}).

\bibitem[{\citenamefont{Henkelman et~al.}(2000)\citenamefont{Henkelman,
  Uberuaga, and J{\'o}nsson}}]{CINEB}
\bibinfo{author}{\bibfnamefont{G.}~\bibnamefont{Henkelman}},
  \bibinfo{author}{\bibfnamefont{B.~P.} \bibnamefont{Uberuaga}},
  \bibnamefont{and}
  \bibinfo{author}{\bibfnamefont{H.}~\bibnamefont{J{\'o}nsson}},
  \bibinfo{journal}{{J}. {C}hem. {P}hys.} \textbf{\bibinfo{volume}{113}},
  \bibinfo{pages}{9901} (\bibinfo{year}{2000}).

\bibitem[{\citenamefont{Henkelman and J{\'o}nsson}(1999)}]{Dimer}
\bibinfo{author}{\bibfnamefont{G.}~\bibnamefont{Henkelman}} \bibnamefont{and}
  \bibinfo{author}{\bibfnamefont{H.}~\bibnamefont{J{\'o}nsson}},
  \bibinfo{journal}{{J}. {C}hem. {P}hys.} \textbf{\bibinfo{volume}{111}},
  \bibinfo{pages}{7010} (\bibinfo{year}{1999}).

\bibitem[{SM()}]{SM}
\bibinfo{note}{See Supplemental Material at [URL will be inserted by publisher]
  for more discussion on the stability of the anchored Pt$_{1}$ species, the
  density of states (DOS) and charge analysis of the PtI$_{4}$ group, the
  structures of the Pt$_{2}$@(110), Pt$_{3}$@(110), and Pt$_{1}$@(001) systems,
  and the effect of terminations on the deposition and catalytic properties of
  the Pt$_{1}$ species.}

\bibitem[{\citenamefont{Yeo et~al.}(1997)\citenamefont{Yeo, Vattuone, and
  King}}]{Yeo1997}
\bibinfo{author}{\bibfnamefont{Y.~Y.} \bibnamefont{Yeo}},
  \bibinfo{author}{\bibfnamefont{L.}~\bibnamefont{Vattuone}}, \bibnamefont{and}
  \bibinfo{author}{\bibfnamefont{D.~A.} \bibnamefont{King}},
  \bibinfo{journal}{J. Chem. Phys.} \textbf{\bibinfo{volume}{106}},
  \bibinfo{pages}{392} (\bibinfo{year}{1997}).

\bibitem[{\citenamefont{Eichler and Hafner}(1997)}]{Eichler1997}
\bibinfo{author}{\bibfnamefont{A.}~\bibnamefont{Eichler}} \bibnamefont{and}
  \bibinfo{author}{\bibfnamefont{J.}~\bibnamefont{Hafner}},
  \bibinfo{journal}{Phys. Rev. Lett.} \textbf{\bibinfo{volume}{79}},
  \bibinfo{pages}{4481} (\bibinfo{year}{1997}).

\bibitem[{\citenamefont{Fu et~al.}(2011)\citenamefont{Fu, Yang, and
  Luo}}]{Fu2011}
\bibinfo{author}{\bibfnamefont{Q.}~\bibnamefont{Fu}},
  \bibinfo{author}{\bibfnamefont{J.~L.} \bibnamefont{Yang}}, \bibnamefont{and}
  \bibinfo{author}{\bibfnamefont{Y.}~\bibnamefont{Luo}}, \bibinfo{journal}{J.
  Phys. Chem. C} \textbf{\bibinfo{volume}{115}}, \bibinfo{pages}{6864}
  (\bibinfo{year}{2011}).

\bibitem[{\citenamefont{Peterson et~al.}(2014)\citenamefont{Peterson, DeLaRiva,
  Lin, Johnson, Guo, Miller, {Hun Kwak}, Peden, Kiefer, Allard
  et~al.}}]{Peterson2014}
\bibinfo{author}{\bibfnamefont{E.~J.} \bibnamefont{Peterson}},
  \bibinfo{author}{\bibfnamefont{A.~T.} \bibnamefont{DeLaRiva}},
  \bibinfo{author}{\bibfnamefont{S.}~\bibnamefont{Lin}},
  \bibinfo{author}{\bibfnamefont{R.~S.} \bibnamefont{Johnson}},
  \bibinfo{author}{\bibfnamefont{H.}~\bibnamefont{Guo}},
  \bibinfo{author}{\bibfnamefont{J.~T.} \bibnamefont{Miller}},
  \bibinfo{author}{\bibfnamefont{J.}~\bibnamefont{{Hun Kwak}}},
  \bibinfo{author}{\bibfnamefont{C.~H.~F.} \bibnamefont{Peden}},
  \bibinfo{author}{\bibfnamefont{B.}~\bibnamefont{Kiefer}},
  \bibinfo{author}{\bibfnamefont{L.~F.} \bibnamefont{Allard}},
  \bibnamefont{et~al.}, \bibinfo{journal}{Nat. Commun.}
  \textbf{\bibinfo{volume}{5}}, \bibinfo{pages}{4885} (\bibinfo{year}{2014}).

\bibitem[{\citenamefont{Liu et~al.}(2016)\citenamefont{Liu, Lucci, Yang, Lee,
  Marcinkowski, Therrien, Williams, Sykes, and
  Flytzani-Stephanopoulos}}]{LiuJL2016}
\bibinfo{author}{\bibfnamefont{J.}~\bibnamefont{Liu}},
  \bibinfo{author}{\bibfnamefont{F.~R.} \bibnamefont{Lucci}},
  \bibinfo{author}{\bibfnamefont{M.}~\bibnamefont{Yang}},
  \bibinfo{author}{\bibfnamefont{S.}~\bibnamefont{Lee}},
  \bibinfo{author}{\bibfnamefont{M.~D.} \bibnamefont{Marcinkowski}},
  \bibinfo{author}{\bibfnamefont{A.~J.} \bibnamefont{Therrien}},
  \bibinfo{author}{\bibfnamefont{C.~T.} \bibnamefont{Williams}},
  \bibinfo{author}{\bibfnamefont{E.~C.~H.} \bibnamefont{Sykes}},
  \bibnamefont{and}
  \bibinfo{author}{\bibfnamefont{M.}~\bibnamefont{Flytzani-Stephanopoulos}},
  \bibinfo{journal}{J. Am. Chem. Soc.} \textbf{\bibinfo{volume}{138}},
  \bibinfo{pages}{6396} (\bibinfo{year}{2016}).

\bibitem[{\citenamefont{N{\o}rskov et~al.}(2014)\citenamefont{N{\o}rskov,
  Studt, Abild-Pedersen, and Bligaard}}]{Norskov-book}
\bibinfo{author}{\bibfnamefont{J.~K.} \bibnamefont{N{\o}rskov}},
  \bibinfo{author}{\bibfnamefont{F.}~\bibnamefont{Studt}},
  \bibinfo{author}{\bibfnamefont{F.}~\bibnamefont{Abild-Pedersen}},
  \bibnamefont{and} \bibinfo{author}{\bibfnamefont{T.}~\bibnamefont{Bligaard}},
  \emph{\bibinfo{title}{{Fundamental Concepts in Heterogeneous Catalysis}}}
  (\bibinfo{publisher}{Wiley}, \bibinfo{year}{2014}), \bibinfo{edition}{1st}
  ed.

\bibitem[{tab()}]{table}
\bibinfo{note}{Http://cccbdb.nist.gov/expvibs1x.asp}.

\bibitem[{\citenamefont{Freund and Roberts}(1996)}]{Freund1996}
\bibinfo{author}{\bibfnamefont{H.-J.} \bibnamefont{Freund}} \bibnamefont{and}
  \bibinfo{author}{\bibfnamefont{M.}~\bibnamefont{Roberts}},
  \bibinfo{journal}{Surf. Sci. Rep.} \textbf{\bibinfo{volume}{25}},
  \bibinfo{pages}{225} (\bibinfo{year}{1996}).

\bibitem[{\citenamefont{Fu and Luo}(2013)}]{FuQ2013}
\bibinfo{author}{\bibfnamefont{Q.}~\bibnamefont{Fu}} \bibnamefont{and}
  \bibinfo{author}{\bibfnamefont{Y.}~\bibnamefont{Luo}}, \bibinfo{journal}{J.
  Phys. Chem. C} \textbf{\bibinfo{volume}{117}}, \bibinfo{pages}{14618}
  (\bibinfo{year}{2013}).

\bibitem[{\citenamefont{Conner and Falconer}(1995)}]{Conner1995}
\bibinfo{author}{\bibfnamefont{W.~C.} \bibnamefont{Conner}} \bibnamefont{and}
  \bibinfo{author}{\bibfnamefont{J.~L.} \bibnamefont{Falconer}},
  \bibinfo{journal}{Chem. Rev.} \textbf{\bibinfo{volume}{95}},
  \bibinfo{pages}{759} (\bibinfo{year}{1995}).

\bibitem[{\citenamefont{Porosoff et~al.}(2016)\citenamefont{Porosoff, Yan, and
  Chen}}]{Porosoff2016}
\bibinfo{author}{\bibfnamefont{M.~D.} \bibnamefont{Porosoff}},
  \bibinfo{author}{\bibfnamefont{B.}~\bibnamefont{Yan}}, \bibnamefont{and}
  \bibinfo{author}{\bibfnamefont{J.~G.} \bibnamefont{Chen}},
  \bibinfo{journal}{Energy Environ. Sci.} \textbf{\bibinfo{volume}{9}},
  \bibinfo{pages}{62} (\bibinfo{year}{2016}).

\bibitem[{\citenamefont{Chang et~al.}(2016)\citenamefont{Chang, Wang, and
  Gong}}]{Chang2016}
\bibinfo{author}{\bibfnamefont{X.}~\bibnamefont{Chang}},
  \bibinfo{author}{\bibfnamefont{T.}~\bibnamefont{Wang}}, \bibnamefont{and}
  \bibinfo{author}{\bibfnamefont{J.}~\bibnamefont{Gong}},
  \bibinfo{journal}{Energy Environ. Sci.} \textbf{\bibinfo{volume}{9}},
  \bibinfo{pages}{2177} (\bibinfo{year}{2016}).

\bibitem[{\citenamefont{Wang et~al.}(2017{\natexlab{a}})\citenamefont{Wang,
  Shi, Li, Chen, and Huang}}]{WangZ2017}
\bibinfo{author}{\bibfnamefont{Z.}~\bibnamefont{Wang}},
  \bibinfo{author}{\bibfnamefont{Z.}~\bibnamefont{Shi}},
  \bibinfo{author}{\bibfnamefont{T.}~\bibnamefont{Li}},
  \bibinfo{author}{\bibfnamefont{Y.}~\bibnamefont{Chen}}, \bibnamefont{and}
  \bibinfo{author}{\bibfnamefont{W.}~\bibnamefont{Huang}},
  \bibinfo{journal}{Angew. Chem. Int. Ed.} \textbf{\bibinfo{volume}{56}},
  \bibinfo{pages}{1190} (\bibinfo{year}{2017}{\natexlab{a}}).

\bibitem[{\citenamefont{Gong et~al.}(2018)\citenamefont{Gong, Yang, Rebollar,
  Rucinski, Liveris, Zhu, and Xu}}]{GongJ2018}
\bibinfo{author}{\bibfnamefont{J.}~\bibnamefont{Gong}},
  \bibinfo{author}{\bibfnamefont{M.}~\bibnamefont{Yang}},
  \bibinfo{author}{\bibfnamefont{D.}~\bibnamefont{Rebollar}},
  \bibinfo{author}{\bibfnamefont{J.}~\bibnamefont{Rucinski}},
  \bibinfo{author}{\bibfnamefont{Z.}~\bibnamefont{Liveris}},
  \bibinfo{author}{\bibfnamefont{K.}~\bibnamefont{Zhu}}, \bibnamefont{and}
  \bibinfo{author}{\bibfnamefont{T.}~\bibnamefont{Xu}}, \bibinfo{journal}{Adv.
  Mater.} \textbf{\bibinfo{volume}{30}}, \bibinfo{pages}{1800973}
  (\bibinfo{year}{2018}).

\bibitem[{\citenamefont{Smith et~al.}(2014)\citenamefont{Smith, Hoke,
  Solis-Ibarra, McGehee, and Karunadasa}}]{Smith2014}
\bibinfo{author}{\bibfnamefont{I.~C.} \bibnamefont{Smith}},
  \bibinfo{author}{\bibfnamefont{E.~T.} \bibnamefont{Hoke}},
  \bibinfo{author}{\bibfnamefont{D.}~\bibnamefont{Solis-Ibarra}},
  \bibinfo{author}{\bibfnamefont{M.~D.} \bibnamefont{McGehee}},
  \bibnamefont{and} \bibinfo{author}{\bibfnamefont{H.~I.}
  \bibnamefont{Karunadasa}}, \bibinfo{journal}{Angew. Chem. Int. Ed.}
  \textbf{\bibinfo{volume}{53}}, \bibinfo{pages}{11232} (\bibinfo{year}{2014}).

\bibitem[{\citenamefont{Zhang et~al.}(2017)\citenamefont{Zhang, Ren, Liu,
  Munir, Zhu, Yang, Li, Liu, Smilgies, Li et~al.}}]{ZhangX2017}
\bibinfo{author}{\bibfnamefont{X.}~\bibnamefont{Zhang}},
  \bibinfo{author}{\bibfnamefont{X.}~\bibnamefont{Ren}},
  \bibinfo{author}{\bibfnamefont{B.}~\bibnamefont{Liu}},
  \bibinfo{author}{\bibfnamefont{R.}~\bibnamefont{Munir}},
  \bibinfo{author}{\bibfnamefont{X.}~\bibnamefont{Zhu}},
  \bibinfo{author}{\bibfnamefont{D.}~\bibnamefont{Yang}},
  \bibinfo{author}{\bibfnamefont{J.}~\bibnamefont{Li}},
  \bibinfo{author}{\bibfnamefont{Y.}~\bibnamefont{Liu}},
  \bibinfo{author}{\bibfnamefont{D.-M.} \bibnamefont{Smilgies}},
  \bibinfo{author}{\bibfnamefont{R.}~\bibnamefont{Li}}, \bibnamefont{et~al.},
  \bibinfo{journal}{Energy Environ. Sci.} \textbf{\bibinfo{volume}{10}},
  \bibinfo{pages}{2095} (\bibinfo{year}{2017}).

\bibitem[{\citenamefont{Li et~al.}(2015)\citenamefont{Li, {Ibrahim Dar}, Yi,
  Luo, Tschumi, Zakeeruddin, Nazeeruddin, Han, and
  Gr{\"{a}}tzel}}]{Nazeeruddin2015}
\bibinfo{author}{\bibfnamefont{X.}~\bibnamefont{Li}},
  \bibinfo{author}{\bibfnamefont{M.}~\bibnamefont{{Ibrahim Dar}}},
  \bibinfo{author}{\bibfnamefont{C.}~\bibnamefont{Yi}},
  \bibinfo{author}{\bibfnamefont{J.}~\bibnamefont{Luo}},
  \bibinfo{author}{\bibfnamefont{M.}~\bibnamefont{Tschumi}},
  \bibinfo{author}{\bibfnamefont{S.~M.} \bibnamefont{Zakeeruddin}},
  \bibinfo{author}{\bibfnamefont{M.~K.} \bibnamefont{Nazeeruddin}},
  \bibinfo{author}{\bibfnamefont{H.}~\bibnamefont{Han}}, \bibnamefont{and}
  \bibinfo{author}{\bibfnamefont{M.}~\bibnamefont{Gr{\"{a}}tzel}},
  \bibinfo{journal}{Nat. Chem.} \textbf{\bibinfo{volume}{7}},
  \bibinfo{pages}{703} (\bibinfo{year}{2015}).

\bibitem[{\citenamefont{Wang et~al.}(2017{\natexlab{b}})\citenamefont{Wang,
  Lin, Chmiel, Sakai, Herz, and Snaith}}]{WangZP2017}
\bibinfo{author}{\bibfnamefont{Z.}~\bibnamefont{Wang}},
  \bibinfo{author}{\bibfnamefont{Q.}~\bibnamefont{Lin}},
  \bibinfo{author}{\bibfnamefont{F.~P.} \bibnamefont{Chmiel}},
  \bibinfo{author}{\bibfnamefont{N.}~\bibnamefont{Sakai}},
  \bibinfo{author}{\bibfnamefont{L.~M.} \bibnamefont{Herz}}, \bibnamefont{and}
  \bibinfo{author}{\bibfnamefont{H.~J.} \bibnamefont{Snaith}},
  \bibinfo{journal}{Nat. Energy} \textbf{\bibinfo{volume}{2}},
  \bibinfo{pages}{17135} (\bibinfo{year}{2017}{\natexlab{b}}).

\end{thebibliography}

\clearpage


\end{document}